\documentclass[letterpaper,journal]{IEEEtran}
\usepackage{amsmath,amsfonts}
\usepackage{algorithmic}
\usepackage{algorithm}
\usepackage{array}
\newcolumntype{P}[1]{>{\centering\arraybackslash}m{#1}}
\usepackage[caption=false,font=normalsize,labelfont=sf,textfont=sf]{subfig}
\usepackage{textcomp}
\usepackage{stfloats}
\usepackage{url}
\usepackage{verbatim}
\usepackage{graphicx}
\usepackage{cite}
\begin{document}

\title{Holovibes : Real-time Ultrahigh-Speed Digital Hologram Rendering and Short-Time Analysis}

\author{
Marius Dubosc, Maxime Boy-Arnould, Jules Guillou, Titouan Gragnic, Arthur Courselle, Gustave Hervé, Alexis Pinson, Etienne Senigout, Bastien Gaulier, Simon Riou,
Chloé Magnier, Noé Topeza, Oscar Morand, Thomas Xu, Samuel Goncalves, Edgar Delaporte, Adrien Langou, Paul Duhot, Julien Nicolle, Sacha Bellier,
David Chemaly, Damien Didier, Philippe Bernet, Eliott Bouhana, Fabien Colmagro, Guillaume Poisson, Anthony Strazzella, Ilan Guenet, Nicolas Blin,
Quentin Kaci, Theo Lepage, Loïc Bellonnet-Mottet, Antoine Martin, François Te, Ellena Davoine, Clement Fang, Danae Marmai, Hugo Verjus,
Eloi Charpentier, Julien Gautier, Florian Lapeyre, Thomas Jarrossay, Alexandre Bartz, Cyril Cêtre, Clement Ledant, Eric Delanghe, Arnaud Gaillard,
Geoffrey Le Gourrierec, Jeffrey Bencteux, Thomas Kostas, Pierre Pagnoux, Antoine Dillée, Romain Cancillière, and Michael Atlan%

\thanks{


This work was supported by the ANR AngioDoppler, ANR VideoLaserDoppler, ERC Synergy HELMHOLTZ, ANR LIDARO, and BPI HoloDoppler grants.


Antoine Dillée and Romain Cancillière are with École Spéciale de Mécanique et d'Électricité ESME Sudria Paris – Centre, 34 rue de Fleurus – 75006 Paris, France.

Thomas Jarrossay, Alexandre Bartz, Clement Ledant, and Eric Delanghe are with 42 – Campus de Paris, 96, boulevard Bessières, 75017 Paris, France.

Marius Dubosc, Maxime Boy-Arnould, Jules Guillou, Titouan Gragnic, Arthur Courselle, Gustave Hervé, Alexis Pinson, Etienne Senigout, Bastien Gaulier, Simon Riou,
Chloé Magnier, Noé Topeza, Oscar Morand, Thomas Xu, Samuel Goncalves, Edgar Delaporte, Adrien Langou, Paul Duhot, Julien Nicolle, Sacha Bellier,
David Chemaly, Damien Didier, Philippe Bernet, Eliott Bouhana, Fabien Colmagro, Guillaume Poisson, Anthony Strazzella, Ilan Guenet, Nicolas Blin,
Quentin Kaci, Theo Lepage, Loïc Bellonnet-Mottet, Antoine Martin, François Te, Ellena Davoine, Clement Fang, Danae Marmai, Hugo Verjus,
Eloi Charpentier, Julien Gautier, Florian Lapeyre, Cyril Cêtre, Arnaud Gaillard,
Geoffrey Le Gourrierec, Jeffrey Bencteux, Thomas Kostas, and Pierre Pagnoux are with the School of Engineering and Computer Science EPITA, 14-16 Rue Voltaire, 94270 Le Kremlin-Bicêtre, France.

Michael Atlan is with the Langevin Institute. Centre National de la Recherche Scientifique (CNRS). Paris Sciences \& Lettres (PSL University). Ecole Supérieure de Physique et de Chimie Industrielles (ESPCI Paris). 1 rue Jussieu, 75005 Paris, France
}
}



\maketitle

\begin{abstract}
Real-time ultrahigh-speed rendering of digital holograms from high-bitrate interferogram streams demands robust parallel computing and efficient data handling with minimal latency. We present Holovibes, a high-performance software engine that enables real-time holographic image reconstruction and short-time analysis at unprecedented throughput. Holovibes integrates spatial demodulation techniques – such as Fresnel transformations and angular spectrum propagation – with temporal analysis methods including short-time Fourier transform (STFT) and principal component analysis (PCA) in a unified pipeline. By leveraging CUDA-based GPU acceleration, multithreaded parallelism, and efficient buffering, the system achieves high-throughput, low-latency processing suitable for demanding computational imaging applications. We demonstrate sustained real-time hologram rendering of 256×256-pixel from interferograms acquired by a streaming camera at 71,400 frames per second on commodity hardware with no frame loss, while maintaining an end-to-end latency of 30 ms. The engine also supports simultaneous recording of raw or processed data, enabling high-speed acquisition workflows essential for experimental applications. This work represents a significant advance over prior digital holography systems and provides a versatile platform for ultra-high-speed, real-time computational imaging.
\end{abstract}

\begin{IEEEkeywords}
real-time, high-speed imaging. computational imaging, Doppler holography, holographic OCT, digital holography, digital holographic microscopy, high performance computing, low latency, GPU.
\end{IEEEkeywords}

\section{Introduction}

\IEEEPARstart{U}{ltrahigh-speed} digital holography represents a transformative advance in coherent light imaging, combining camera sensor arrays with digital signal processing methods inspired by radar technology. These approaches enable low-light imaging, phase-resolved optical wavefield measurements, and post-acquisition wavefield reconstruction and manipulation~\cite{kumar2013subaperture, javidi2021roadmap, Hillmann2016, rosen2024roadmap}. The increasing availability of ultrafast cameras and high-throughput GPU computing has triggered the emergence of real-time digital holographic techniques that will predictably drive the democratization of compute-intensive imaging in real-time ~\cite{Leutenegger2011Real, Puyo2020Realtime, NAKAI2022, nagahama2025introducing, morales2025holostream}. 

In medical imaging, digital holography already enables high-quality, label-free structural and functional imaging of the human retina in offline settings~\cite{ Hillmann2016, Puyo2018Vivo, tomczewski2022light}. It is poised to advance ophthalmic modalities such as OCT and Doppler angiography, but real-time implementation remains limited by acquisition and processing demands. Swept-source holographic OCT of the human retina requires recording of about 600 frames within 10 ms to preserve optical phase stability~\cite{hillmann2012common, spahr2019phase}, while Doppler holography of retinal blood flow demands similarly high frame rates for accurate velocity and flow quantification~\cite{Puyo2020Spatio}.

Ultrafast volumetric methods, such as holographic optoretinography, reveal stimulus-evoked phase shifts in photoreceptors~\cite{spahr2019phase, tomczewski2022light} and ganglion cells~\cite{pfaffle2022phase}, and now support wide-field, high-resolution imaging. Real-time deployment of these techniques hinges on ultrafast cameras and GPU pipelines capable of sustaining gigavoxel-scale data rates. Likewise, Doppler holography overcomes historical limitations in retinal blood flow quantification, providing full-field, high-temporal-resolution maps of pulsatile flow in retinal vessels previously inaccessible to OCT or angiography~\cite{puyo_vivo_2018}. This enables the extraction of quantitative hemodynamic biomarkers critical for assessing ocular and systemic vascular health~\cite{fischer_retinal_2024}.

Early efforts to harness GPU acceleration for digital holography led to the development of the GWO library~\cite{shimobaba2008numerical}, enabling real-time digital holographic microscopy~\cite{shimobaba2008real} and multi-view, multi-resolution visualization~\cite{shimobaba2010real}. Aimed at optics researchers with limited experience in general-purpose GPU programming, GWO provided basic GPU-accelerated diffraction and CGH computations. However, it required manual memory management and was limited to GPU-only execution, reducing its accessibility and flexibility. An upgraded version of the routines, the CWO++ library, was introduced as a more user-friendly C++ framework supporting both CPU and GPU computation for diffraction~\cite{shimobaba2012computational}. This evolution reflects a broader trend toward high-performance, developer-accessible holographic rendering frameworks compatible with parallel computing architectures.

Our previous real-time digital holography software for holographic OCT and Doppler holography were limited to a few hundred frames per second~\cite{charpentier2020swept, puyo_real-time_2020}, were proved insufficient to meet the throughput demands of these applications. Accurate measurement of broadband optical fluctuations, central to functional and flow imaging, requires acquisition rates of the order of tens of thousands of frames per second and low-latency pipelines capable of sustaining this bandwidth with zero frame loss.

Similar high-throughput demands are emerging in other domains. In fluid dynamics, digital in-line holography at 20,000 fps captures transient multiphase flow phenomena~\cite{guildenbecher2016high}. In industrial metrology and materials science, ultrafast digital holographic interferometry enables full-field vibration and motion analysis, often requiring frame rates exceeding 100,000 fps~\cite{kakue2017digital}. Despite their diversity, these applications share a common need for flexible, high-performance imaging systems capable of processing massive data rates in real time. Until now, most ultrafast holography demonstrations have relied on offline reconstruction or specialized hardware, as general-purpose platforms struggled to balance throughput and latency in configurable software environments.

Hardware-accelerated rendering techniques have recently drawn attention. Field-programmable gate arrays (FPGAs) can execute Fresnel diffraction calculations up to 23 times faster than GPUs for small-frame inputs~\cite{yamamoto2020special, hara2022design}, making them highly efficient for embedded or productized imaging systems. However, FPGAs lack the flexibility of general-purpose GPUs, which are better suited for algorithmic development and complex multi-threaded pipelines in scientific applications.

To address these constraints, we developed Holovibes~\cite{Holovibes} —an open-source, high-throughput interferogram rendering engine for real-time processing, visualization, and recording. Holovibes integrates spatial demodulation techniques (e.g., Fresnel and angular spectrum propagation) with fast temporal analysis methods (e.g., short-time Fourier transform and PCA) into a unified, GPU-accelerated pipeline. The architecture decouples acquisition, processing, and rendering using lock-free ring buffers in GPU memory, ensuring asynchronous operation without frame drops. This design enables seamless streaming of interferograms at extreme data rates with minimal latency, leveraging CUDA for efficient GPU memory transfer and processing.

In benchmarking tests, Holovibes achieved sustained holographic rendering at 71,400 frames per second on a single commodity GPU, with no frame loss. This performance far exceeds previous real-time systems and opens new possibilities for high-speed functional and flow imaging in biomedical applications, as well as broader use cases in scientific and industrial settings.

The contributions of this work include :
\begin{itemize}
    \item A novel real-time holographic imaging architecture: We design a multi-threaded, GPU-accelerated pipeline for high-throughput, low-latency hologram rendering and analysis of streaming interferograms. This architecture robustly renders and records tens of thousands of frames per second without dropping data by using parallel tasks and buffer management.
    We demonstrate sustained hologram rendering at 71,400 fps with about 30 ms latency on a single GPU workstation, with no frame loss or interruptions.
    \item Integration of spatial and short-time temporal analyses: Our system seamlessly combines off-axis hologram reconstruction (angular spectrum/Fresnel propagation) with short-time Fourier transform processing for Doppler analysis, as well as other computational tools like PCA, all in real-time. This integration enables simultaneous visualization and quantitative analysis (e.g. blood flow estimation) on live data streams.
    \item Versatility and open-source implementation: Holovibes is device- and application-agnostic – it can ingest data from ultra-high-speed cameras or file streams and output holographic images and spectra in real-time. The software has been made available as an open-source C++/CUDA project.
\end{itemize}
Through these contributions, we aim to bridge the gap between cutting-edge high-speed imaging hardware and the computational frameworks required to fully utilize their data in real time.

\section{Image rendering engine}

\subsection{Specification}
The requirement of the image rendering engine is twofold: 
\begin{itemize}
    \item To acquire and record a continuous, high-bitrate digital stream of optically captured interferograms from an ultra-high-speed streaming camera.
    \item To process this interferogram stream with low latency, rendering numerical holograms for real-time visualization. 
\end{itemize}

In ultrahigh-speed digital holography, recording raw frames for digital processing requires managing high data throughput, as the captured interferograms are essential for subsequent image reconstruction. Any data loss, such as dropped frames, compromises the integrity of short-time fluctuations in coherent light, potentially distorting the reconstructed images.
Real-time visualization demands hologram rendering with minimal latency while simultaneously handling the camera’s high data throughput. Although frame loss is less critical for visualization, both recording and visualization must operate in parallel during acquisition to ensure continuous and reliable data capture.
Real-time digital hologram rendering at high frame rates therefore requires high data bitrate, avoidance of frame drops in both input and recorded data streams, and minimal latency between acquisition and display. These combined requirements—high throughput, full data integrity, and low latency—necessitate a specialized digital hologram rendering engine. This engine must independently perform spatial and temporal signal demodulation, ensuring versatility across computational imaging applications while sustaining high performance.

\subsection{Solutions}
The software design addressing the requirements of real-time, high-bitrate digital hologram rendering relies on bufferization and parallelization. An end-to-end buffer management and asynchronous queue system is implemented for congestion control. This system efficiently performs spatial and temporal transformations on buffered data, asynchronously from the high-frame-rate camera acquisition and output stream management.
The solution implements multiple forms of parallelization: acquisition, rendering, and recording tasks are executed in parallel; computations are offloaded to the GPU; and data transfers between CPU and GPU occur concurrently with ongoing computations.
Together, these techniques form a robust and efficient pipeline capable of meeting the high-throughput, low-latency, and data-integrity demands of real-time digital holography. By optimizing resource usage and balancing workloads across the CPU and GPU, this design ensures a responsive, high-performance rendering engine adaptable to diverse computational imaging scenarios.

\subsubsection{Buffers and queues}

Bufferization, which consists of temporarily storing data in memory buffers, is crucial for streamlined computations. Two types of buffers are used in Holovibes :
\begin{itemize}
    \item Fixed-size buffers used for synchronous operations applied to the entire buffer within a single execution thread (e.g., spatial and temporal transformations). These buffers are precisely sized according to the data type and the user-defined frame count for each operation, simplifying and optimizing the computation pipeline.
    \item Queue buffers for asynchronous access, used as intermediaries between a producer (e.g., the computation pipeline generating rendered images) and a consumer (e.g., the recording thread processing these images for storage). These ring FIFO (First-In, First-Out) buffers support asynchronous access and are thread-safe, allowing the producer and consumer to operate at independent rates without interference. Each queue buffer has a defined capacity to manage data flow and prevent overflow. To minimize overhead and locking, the input queue (which bridges the camera and the computation pipeline) uses a packet-based locking approach: frames are enqueued and dequeued in packets, enabling concurrent access by producer and consumer threads.
\end{itemize}

\subsubsection{Task parallelization}

To enable concurrent execution, each task is assigned to a separate execution thread. This approach ensures efficient use of computational resources, minimizes processing delays, and allows each task to progress asynchronously.
The main execution threads, referred to as workers, are: the FrameReadWorker, which handles image acquisition; the ComputeWorker, which manages the computation pipeline, including all operations applied to the frames; and the RecordWorker, which is responsible for recording the frames.
The master thread of the program, managed by the Qt interface, creates and controls all worker threads and is responsible for displaying the processed images.

\subsubsection{Computation parallelization}
The engine sequentially conducts spatio-temporal transformations by parallel processing on a GPU. We make use of the GPU through CUDA (Compute Unified Device Architecture), a parallel computing platform that gives direct access to the GPU's virtual instruction set and parallel computational elements for the execution of compute kernels.
The essential computing operations for hologram rendering include spatial Fourier transformations, used either for angular spectrum propagation or Fresnel transformation, and temporal transformations on sequences of reconstructed
holograms, namely short-time Fourier transformation (STFT), principal component analysis (PCA), singular spectrum analysis (SSA), or a combination of these methods.

\subsubsection{Data transfer parallelization}

To maximize GPU efficiency, we use CUDA streams to separate computations from memory transfers, allowing parallel execution of data reading, writing, and tasks processing.
Unlike sequential processing, where tasks run one after another (Table \ref{tab:sequential}), CUDA streams allow data transfers and computations to occur concurrently, leveraging the GPU’s capacity for simultaneous operations (Table \ref{tab:concurrent_packet}). In our approach, each stream is dedicated to a specific operation: inserting data (Host to Device memory copy), processing data (Kernel execution), and retrieving data (Device to Host memory copy) (Table \ref{tab:concurrent_stream}). Although the tables illustrate this process with three frame packets for simplicity, in practice, we handle many more, subdividing the processing tasks into multiple streams to fully exploit the computational power of the GPU.

\setlength{\tabcolsep}{3pt} 
\renewcommand{\arraystretch}{1.6} 
\begin{table}[ht]
\centering
\caption{Sequential Processing of a single frame packet. At each time, only one operation is performed because only one frame packet is being processed.} 
\label{tab:sequential}
\begin{tabular}{|P{13mm}|P{20mm}|P{20mm}|P{20mm}|}
\hline
 & time 1 & time 2 & time 3\\
\hline
frame packet 1 & insert (insert frame packet into input queue) & process (hologram rendering and temporal demodulation) & retrieve (retrieve processed images from output queue) \\
\hline
\end{tabular}
\end{table}

\setlength{\tabcolsep}{3pt} 
\renewcommand{\arraystretch}{1.6} 
\begin{table}[ht]
\centering
\caption{Concurrent Processing of multiple frame packets. At each time, up to three operations can be performed concurrently because frame packets can be inserted or processed or retrieved in parallel.}
\label{tab:concurrent_packet}
\begin{tabular}{|P{13mm}|P{10mm}|P{10mm}|P{10mm}|P{10mm}|P{10mm}|P{10mm}|}
\hline
 & time 1 & time 2 & time 3 & time 4 & time 5 & time 6\\
\hline
frame packet 1 & insert & process & retrieve & &  & \\
\hline
frame packet 2 &  & insert & process & retrieve & & \\
\hline
frame packet 3 & & & insert & process & retrieve &  \\
\hline
frame packet 4 & & & & insert & process & retrieve \\
\hline
\end{tabular}
\end{table}

\setlength{\tabcolsep}{3pt} 
\renewcommand{\arraystretch}{1.6} 
\begin{table}[ht]
\centering
\caption{Concurrent processing via CUDA streams. Parallel processing of compute.} 
\label{tab:concurrent_stream}
\begin{tabular}{|P{13mm}|P{10mm}|P{10mm}|P{10mm}|P{10mm}|P{10mm}|P{10mm}|}
\hline
 & time 1 & time 2 & time 3 & time 4 & time 5 & time 6\\
\hline
CUDA stream 1 & insert & insert & insert & insert &  & \\
\hline
CUDA stream 2 &  & process & process & process & process & \\
\hline
CUDA stream 3 & & & retrieve & retrieve & retrieve & retrieve \\
\hline
\end{tabular}
\end{table}

\section{Data workflow}

\subsection{Input data stream}
Images acquired from a camera or a file are buffered to avoid frame drop in a GPU RAM ring buffer: the input queue. The frames are transferred in packets, to mitigate the latency of memory copy operations. 
To enqueue these frame packets in the GPU-resident input queue, a mapped pinned memory buffer is allocated on the CPU and made accessible to the GPU. Pinned memory serves as a staging area for transfers between the device and host, increasing data throughput and reducing potential delays by bypassing the overhead of transfers between pageable and pinned GPU memory. In this way, raw camera data is continuously written to the input queue, packet by packet, allowing rapid data transfer from CPU to GPU memory.
The input queue functions as a thread-safe producer-consumer queue, where the FrameReadWorker enqueues frame packets from the camera and the ComputeWorker dequeues frame batches for processing in the computation pipeline. Frame packets, optimized for data transfer from the frame grabber, must be distinguished from frame batches, the basic units for spatial and temporal processing.

\subsection{Computation pipeline}
The main component of the program is the computation pipeline, called the pipe, which aims to streamline operations for image rendering. The pipe is a dynamic array of functions that are applied sequentially and repeatedly to the input data stream (frames) from a file or camera. In practice, the computation pipeline consists of a set of parameterised deterministic functions such as discrete Fourier transformations. When a user modifies a reconstruction parameter for image rendering, the pipeline is re-created anew, in order to update the computations accordingly to the new settings. These settings include hologram reconstruction distance, choice of wave propagation algorithm, type of temporal demodulation, short-time analysis window size for temporal signal demodulation, or post-processing options. All computations within the pipeline are done synchronously, but the pipeline runs asynchronously with respect to input (the data stream from the camera does not stop) and output (image display and recording operate independently).

\begin{figure}[!t]
\centering
\includegraphics[width=\linewidth]{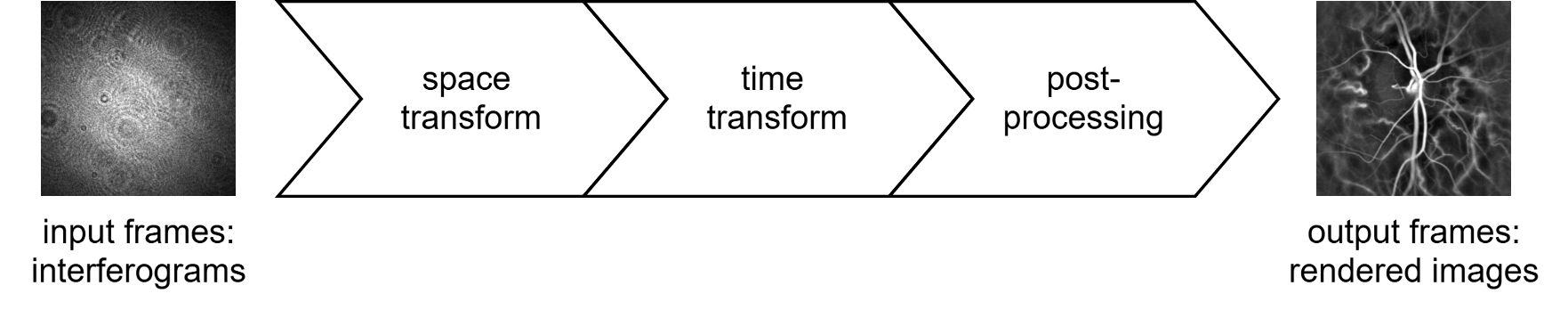}
\caption{Processing pipeline. Two-step computation workflow for real-time dynamic holographic imaging. Spatial Transformations render holograms from each buffered frame, and Short-Time Temporal Transformations to capture fine temporal changes within short intervals. Then, post-processing operations like convolutions or registration can be applied, and a final accumulation sharpens the resulting images.}
\label{pipeline}
\end{figure}

\subsubsection{Space and time transformations}
The processing pipeline is summarized in Fig. \ref{pipeline}. Each batch of frame undergoes a two-step parallel processing: 1- a spatial transformation to create a reconstructed hologram from each recorded interferogram, and 2- a temporal transformation applied to a given number of consecutive holograms.
Whenever the input queue contains one full batch of frames, input data is converted to complex representations and copied to the space transformation buffer frame batch by frame batch. This buffer is used to apply a 2D spatial Fast Fourier Transform (FFT) to the extracted batch. This operation is applied in parallel to each frame of the batch with cuFFT using XtPlanMany data shaping. Applying the transformation to a batch of images allows to reduce the number of memory copies and increase the processing speed. This operation overwrites the space transformation buffer.
Then, the batch of holograms is moved into the time transformation queue of arbitrary user-selected size. This data structure is filled with holograms frame batch by frame batch.
The short-time temporal transform of holograms in the time transformation queue can be either done by FFT, Principal Component Analysis (PCA), or Singular Spectrum Analysis (SSA) and FFT, where the underlying SVD are achieved by data eigen-decomposition. The construction of a covariance matrix of batches of holograms, its eigen-decomposition, and the projection of the result into a sub-basis are performed with cublasCgemm and cusolverDnCheevd.
Then, the magnitude, argument, or statistical moment at a given pixel, can be calculated over a given range of indices within the time transformation buffer.

\subsubsection{Post processing and image accumulation}
The computed magnitude, argument or statistical moment resulting from the temporal transformation ends up in a buffer dedicated to post processing operations, such as image convolution, image re-normalization, image registration and contrast correction.
The convolution is performed on the image with given convolutions kernels such as gaussian filters, or with user designed kernels. The re-normalization is performed after the convolution.
The image registration is performed with a cross-correlation between incoming post process frames and a reference frame. The reference frame is obtained by accumulating frames and computing their mean within a circular mask. Each rendered image is then cross-correlated with this reference frame, and shifted by the offset given by the peak of the cross-correlation.
This new frame data is copied to an image accumulation buffer. This buffer of arbitrary user defined size performs a moving average operation. The buffer is filled and emptied frame by frame, concurrently. By accumulating frames in such a way, we get a sharper image.

\subsection{Output Data Stream}
Once a frame has been post-processed, it is moved to the record queue for record, and/or the output queue for display. Those queues are used to circumvent latency and throughput issues during image recording and display, and to make the computation pipeline asynchronous to memory copy towards the CPU, handled by other threads (the main thread for display, and the RecordWorker for the record). This asynchronous buffering system enables the entire pipeline to process input and output streams concurrently.
Both raw camera frames and rendered images can be recorded. Recording the raw data enables further offline computations, but comes with a high memory cost. For raw record, the frames are copied directly from the input queue in the record queue. For hologram record, the hologram frames rendered via computations are copied from the last buffer of the pipe to the record queue. Depending on the type of record (hologram or raw), the output can be exported respectively as a video .mp4 or .avi or as a custom .holo file which contains raw data and image reconstruction parameters.
When a frame has been successfully rendered or recorded, a new frame is retrieved from the queue and copied to the CPU memory to be displayed and/or recorded on the disk.

\begin{figure*}[!t]
\centering
\includegraphics[width=0.9\linewidth]{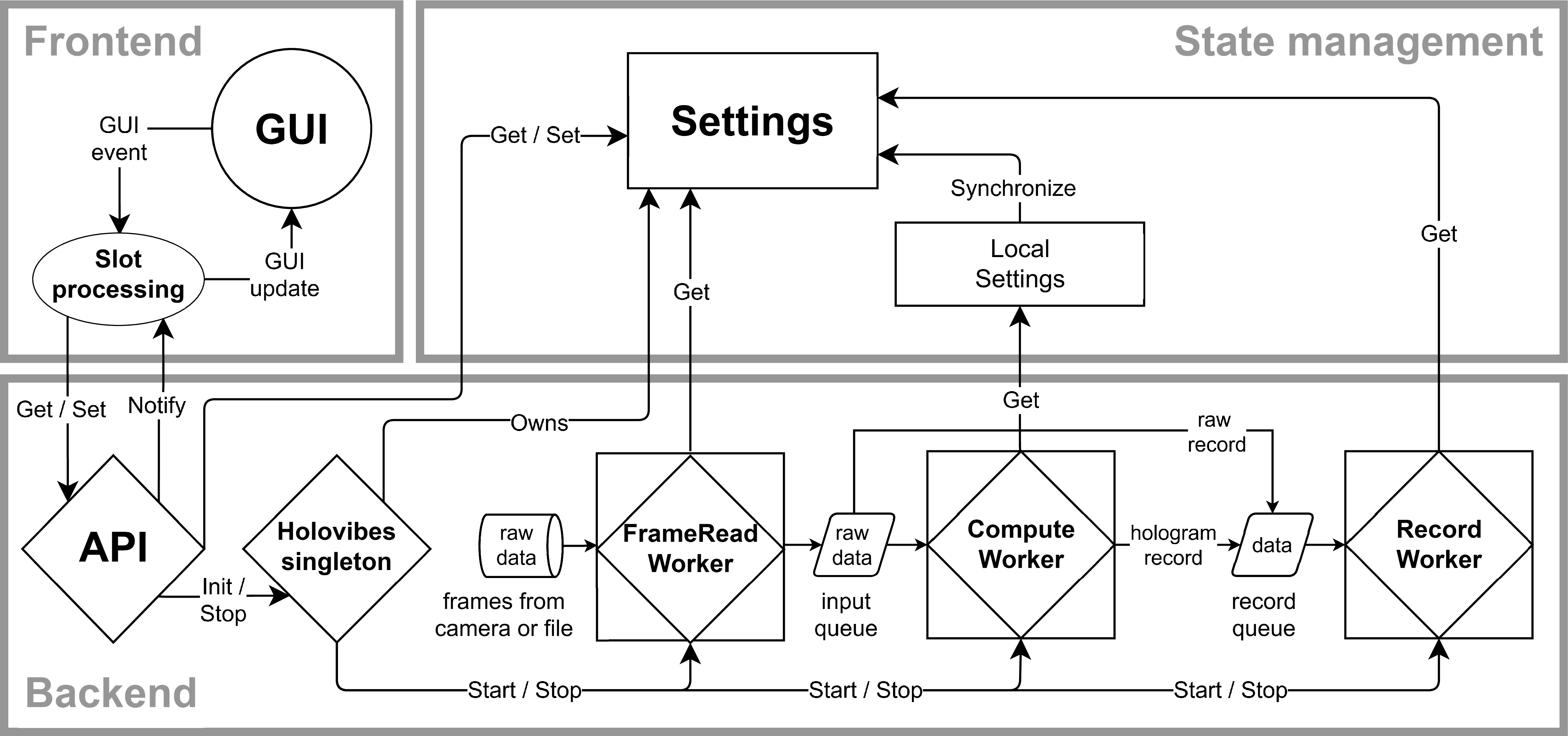}
\caption{Flowchart of the architecture of Holovibes, structured into Front-End, Back-End, and State Management components. The Front-End provides a GUI and CLI for user interaction. The Back-End houses the API, managing program variables through getters and setters and defining the computation pipeline. Leveraging a multithreaded architecture, Holovibes distributes tasks across the FrameReadWorker, RecordWorker and ComputeWorker. Rules surround the state management : the global settings are owned by the Holovibes singleton, and modified only through the API. The pipeline keeps a local version of some of those settings, so that it can run while state modification occur}
\label{archi}
\end{figure*}

\section{Software architecture}
As depicted in Fig. \ref{archi}, Holovibes is structured into three main components:
\begin{enumerate}
    \item Back-End: Processes the frames through its computation pipeline, and contains the API (Application Programming Interface), which manages getters and setters for all program variables which define the computation pipeline.
    \item Front-End: Comprising a Graphical User Interface (GUI) and Command Line Interface (CLI), allowing users to interact with the program.
    \item State Management: Management of all program parameters, including compute settings and user configurations.
\end{enumerate}

\subsection{Front-End}
The Holovibes GUI (Graphical User Interface) front-end provides users with an interactive platform to control data acquisition, processing, and visualization. Built with Qt and optimized for multithreading, the front-end allows users to load cameras or files, adjust acquisition settings, modify compute parameters, and define output formats. A real-time display window shows the results of hologram processing, with immediate updates reflecting changes to environment variables for rapid, visual parameter tuning.
Alternatively, Holovibes can be operated through a Command Line Interface (CLI). In CLI mode, the user provides a file with raw data, an optional configuration file specifying compute settings, and an output path for storing results.
The CLI is typically used for delayed processing of recorded interferograms, particularly when operations are too computationally intensive for real-time rendering. This delayed processing can also be performed in the GUI by loading a file instead of connecting to a camera, with the added benefit of interactively adjusting rendering parameters.
The CLI is especially useful for automated batch processing, allowing users to script the processing of large numbers of input files with predefined compute settings. A PowerShell script has been developed and made publicly available for this purpose, along with preset configuration files tailored to specific imaging applications.
Additionally, the CLI is used to perform non-regression and functional tests.
\subsection{State management}
\begin{figure*}[!t]
\centering
\includegraphics[width=0.9\linewidth]{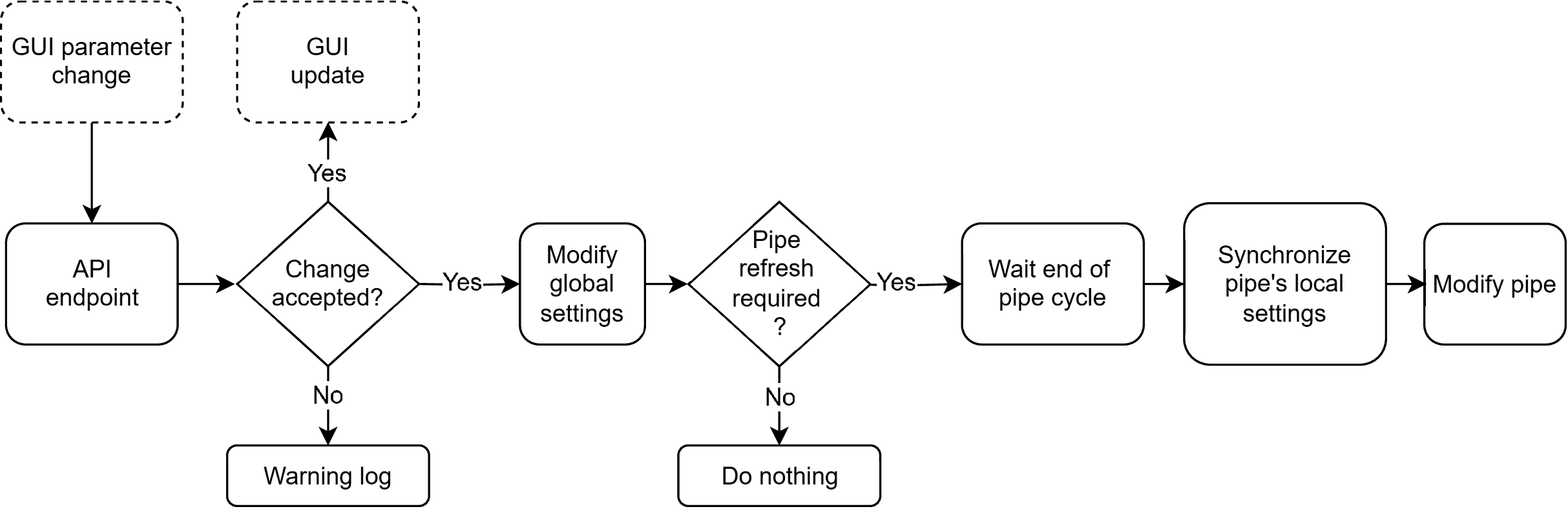}
\caption{Flowchart of the modification of a setting. Frontend logic is represented in dotted lines, backend logic in continuous lines. The user changes a parameter on the GUI. The API processes the modification, handling all backend logic. If the modification has met its preconditions, the global setting is modified by the API. If the setting is a compute setting used within the pipe and/or whose modification affects the pipe, a refresh is requested. At the end of the pipe cycle (when its current batch is processed), the pipe synchronizes its local settings with the global settings, and is cleared then rebuilt accordingly to the new setting.}
\label{state}
\end{figure*}

A multithreaded software often uses a data structure or object that stores the state of the entire program, and is accessible to all threads. It is used to coordinate the activities of multiple threads and ensure that they are all operating with the same parameters. The Holovibes settings, designed to be shared between the back-end and the front-end, are stored by the Holovibes singleton handled by the main thread (Fig. \ref{archi}). They are made accessible to all threads in the program through the API, which provides methods for reading and writing the state of the program while ensuring that the state is consistent, even when multiple threads are accessing it. The settings’ modification process is depicted in Fig. \ref{state}.
Parameters affecting the computation pipeline, such as the choice of spatial and temporal transformations, are called compute settings. The computation pipeline of the backend, handled by the ComputeWorker, needs to keep local versions of those settings, since an update occurring in the middle of the computation pipeline would corrupt the results or lead to segmentation faults. There are therefore two types of settings, depending on their synchronization rules :
\begin{itemize}
    \item \textit{Realtime} settings: settings that can be always synchronized with the global state, because they either do not affect the computation pipeline, or are only used to build the pipeline, and not used during its execution (like the choice of spatial transformation) 
    \item \textit{Delayed} settings: local settings used within the pipeline (Fig. \ref{archi}), needing the current computations to finish before being updated (like batch size or time stride). They are updated by synchronization with the global settings (Fig. \ref{state})
\end{itemize}

\subsection{Back-End}
The back-end of Holovibes leverages a multithreaded architecture to manage frame reading, computations, and frame recording in parallel. By distributing these tasks across multiple threads, Holovibes significantly enhances computational efficiency, delivering high-throughput image rendering with low latency. Instruction parallelization occurs at both the CPU and GPU levels: CPU tasks are handled by worker threads, while GPU computations are managed by CUDA kernels within the ComputeWorker. The main workers are the following:
\begin{itemize}
    \item FrameReadWorker: Manages memory allocation and data transfer from the CPU (host) to the GPU (device).
    \item ComputeWorker: Executes parallelized spatial and temporal transformations and post-processing tasks on GPU CUDA cores (streaming multiprocessors) through CUDA kernels.
    \item RecordWorker: Handles memory allocation and data transfer from the GPU back to the CPU.
\end{itemize}

The API (Application Programming Interface) connects the front-end to the back-end, by processing the user’s updates, modifying the settings accordingly, and triggering actions such as pipe re-updating, frame record, file load, etc. Each front-end function, called slot, has a corresponding function in the API, working as an endpoint, handling all back-end logic and state modification, leaving front-end modification to the front-end slots. 
The API is responsible for setting rules around the flow of variables and settings from the front-end to the back-end. The access and modification of the state is done by API queries and commands, in order to centralize the state management and make it independent from the front-end. This facilitates development and ensures better scalability and maintainability.
A singleton pattern is used to launch execution threads from the software API, providing a single, coordinated access point to manage shared resources and settings in Holovibes.

\section{Holofile system}
The .holo files loaded by Holovibes are divided in three parts : the header, which contains the metadata of the file such as the dimensions and number of images, the raw data, composed of the images, and the footer, containing the image rendering parameters (compute settings), used during the recording of the file.

\begin{itemize}
    \item Header syntax 64-byte binary:
    \begin{itemize}
    \item "HOLO" magic number (4 bytes)
    \item Version number : 7 (2 bytes)
    \item Number of bits per pixel (2 bytes)
    \item Width of the images (4 bytes)
    \item Height of the images (4 bytes)
    \item Number of images (4 bytes)
    \item Total data size in bytes (8 bytes)
    \item Endianness (1 byte) 0/1 little/big endian
    \item Data type (1 byte) 0 raw, 1 processed
    \item Padding to make the header 64 bytes long
    \end{itemize}
    \item Raw image data is written and read as a binary stream of data of length: the number of bits per pixel has to be a multiple of 8 since Holovibes only reads an integer number of bytes per pixel (from files).
    \item The footer data is optional. It contains the image rendering parameters that were used for visualization during the recording of the raw data. It improves the user experience by restoring those image rendering parameters when loading the file in Holovibes. It therefore enables the saving of a given rendering while keeping the possibility to modify or reset the rendering. The data is a collection of key value fields in JSON format. Some fields may be unused while required fields will be filled with sensible defaults values if missing. The different versions of the .holo format and their compute settings are documented in a public document, available on a dedicated website \cite{holofile}.
\end{itemize}

\section{Benchmarks}

\subsection{Hardware and software configuration}
We evaluate the performance of digital hologram rendering using Holovibes, leveraging the CUDA 12 toolkit on a Microsoft Windows 11 64-bit system. This experimental demonstration was enabled through the implementation of the software version 14.8.3 described in this report. The test computer featured commodity hardware components (CPU: Intel Xeon w7-2475X, Storage: Samsung 990 PRO M.2 SSD, 4TB, Motherboard: ASUS Pro WS W790-ACE, RAM: 64 GB Samsung DDR5-4800 Mhz RDIMM). Performance benchmarks included continuous acquisition and/or rendering of Doppler holograms, powered by an NVIDIA GeForce RTX 4090 GPU. These tests involved processing optical interferograms in real-time, acquired via an Ametek Phantom S711 high-speed streaming camera, through two Euresys Coaxlink QSFP+ CoaXPress frame grabbers.

\subsection{Performance analysis}

\subsubsection{Angular Spectrum Propagation with STFT (512-frame window, 512-frame stride)}

This benchmark is intended for 3D image rendering from swept-source holographic OCT setups and instruments requiring equivalent processing pipelines. Holograms were reconstructed using the angular spectrum propagation method, and a high pass spatial filter was applied. Temporal demodulation was achieved via short-time Fourier transform (STFT) applied over 512-frame windows, with a stride of 512 frames. The following sustained performance levels were observed during camera-based acquisition (Tab. \ref{tab:benchmarks}):

\begin{itemize}
    \item 1280 $\times$ 800 pixels: 5,780 frames per second, corresponding to 5.9 Gigavoxels/s.
    \item 512 $\times$ 512 pixels: 22,700 frames per second, yielding 5.9 Gigavoxels/s.
    \item 512 $\times$ 320 pixels: 35,700 frames per second, resulting in 5.8 Gigavoxels/s.
    \item 384 $\times$ 384 pixels (camera-limited): 38,500 frames per second, corresponding to 5.7 Gigavoxels/s.
    \item 256 $\times$ 256 pixels: 71,400 frames per second, yielding 4.7 Gigavoxels/s.
\end{itemize}

\subsubsection{Fresnel Transform with PCA (32-frame window, 32-frame stride)}

This benchmark is intended for image rendering from Doppler imaging setups and instruments requiring equivalent processing pipelines. Hologram rendering was performed using the Fresnel transform. Temporal demodulation employed principal component analysis (PCA) \cite{puyo_real-time_2020} over 32-frame windows, repeated every 32 frames. The following sustained performance levels with minimal latency were achieved during camera-based acquisition (Tab. \ref{tab:benchmarks}):

\begin{itemize}
    \item 1280 $\times$ 800 pixels (camera-limited): 7,360 frames per second, corresponding to 7.5 Gigapixels/s.
    \item 512 $\times$ 512 pixels (camera-limited): 23,578 frames per second, yielding 6.2 Gigapixels/s.
    \item 512 $\times$ 320 pixels: 31,200 frames per second, with a throughput of 5.1 Gigapixels/s.
    \item 384 $\times$ 384 pixels: 33,882 frames per second, corresponding to 5.0 Gigapixels/s.
    \item 256 $\times$ 256 pixels: 58,800 frames per second, giving a rate of 3.9 Gigapixels/s.
\end{itemize}

\setlength{\tabcolsep}{3pt} 
\renewcommand{\arraystretch}{1.6} 
\begin{table}[ht]
\centering
\caption{Performance analysis of different computation pipelines with holovibes v14.8.3. file loaded in GPU VRAM.}
\label{tab:benchmarks}
\begin{tabular}{|P{15mm}|P{15mm}|P{15mm}|P{15mm}|P{15mm}|}
\hline
spatial transform & \multicolumn{2}{|P{30mm}|}{angular spectrum propagation \& spatial filtering} & \multicolumn{2}{|P{30mm}|}{Fresnel transform} \\
\hline
time transform & \multicolumn{2}{|c|}{STFT 512 frames} & \multicolumn{2}{|c|}{PCA 32 frames} \\
\hline
time stride &
\multicolumn{2}{|c|}{512 frames} & \multicolumn{2}{|c|}{32 frames} \\
\hline
input
frame (pixels) & file
(FPS)
 & camera
(FPS)
 & file
(FPS)
 & camera
(FPS) \\
\hline
1280 $\times$ 800 & 6,100 & 5,780 & 6,650 & 7,360 camera-limited \\
\hline
512 $\times$ 512 & 23,100 & 22,700 & 25,200 & 23,578 camera-limited \\
\hline
512 $\times$ 320 & 35,900 & 35,700 & 37,850 & 31,200 \\
\hline
384 $\times$ 384 & 39,000 & 38,500 camera-limited & 40,270 & 33,882 \\
\hline
256 $\times$ 256 & 88,100 & 71,400 & 53,090 & 58,800 \\
\hline
\end{tabular}
\end{table}

\section{Discussion}

\subsection{API as a library}
Ideally, an API is completely agnostic to the front-end ; our goal is to achieve this agnosticism, and make the back-end available as a library, compatible with any front-end as long as it connects with the API endpoints.
For now, the QT front-end cannot be separated from the backend. Most of the logic has been correctly entrusted to the API, and the front-end can no longer affect the state without calling the API, but the front-end still uses some custom structures and backend files that either should be dissociated from the backend, or accessed through the API.
Furthermore, the project cannot be compiled and run without QT, which manages its main thread.
The separation of front-end and backend logic and the circumscription of state management have finally reached a point allowing the complete independence of the backend, pilotable through the API. Future developments will be dedicated to this separation.

\subsection{Dynamic computation pipeline}
The computation pipeline (the pipe) currently consists of a vector of functions (mainly CUDA kernels) executed sequentially. It relies on predefined queues and buffers, and local settings with different synchronization rules. The functions are “statically” inserted within the `refresh()` method during the pipe’s construction. Their insertion depends on settings evaluated at run time, but the design of the pipe is static, since the order of insertions is known at compile time. This design makes the code harder to understand and modify, and has several flaws:
\begin{itemize}
    \item The operations and their order are fixed. It makes the code poorly scalable, hinders versatility and prevents any plugin possibility.
    \item It requires the use of countless booleans, which weighs down the code, as well as the use of anonymous functions (lambda functions) inserted in the pipe’s function vector, which make the code understanding, modification and debugging really difficult. This design also relies on specific buffers for each operation, which is a non optimal memory allocation strategy.
\end{itemize}

A main lead of improvement would be to adopt a dynamically oriented design, in which the computation pipeline would take the form of a computational graph, which would circumvent those flaws, and allow to freely chain operations as long as the data dimensions match.
This dynamic design, combined with the API as a library, would allow the user to access the computation pipeline through a dynamic C++ library or python endpoints, to combine operations freely and even add custom kernels, in order to design his own custom computation pipeline. It would render Holovibes more flexible and scalable, as well as easier to maintain and understand for contributing developers.

\subsection{Minimizing data allocations and transfers}
Handling memory efficiently is crucial when designing software based on an intensive GPU computation pipeline. We used NVIDIA Nsight Systems to analyze GPU memory utilization and avoid memory re-allocations. As GPU memory allocation is very costly, we allocate all buffers only once when the program starts and free them when the program stops. The re-allocations happen when the user changes the size of the asynchronous queues or the size of the batches used in the pipe (batch size, time transformation size). Similarly, memory transfers between the CPU and the GPU or between buffers were reduced to a strict minimum to limit the number of calls to cudaMalloc and cudaMemcpy. Since the number of frames needed for each computation is always known, the memory usage could be further optimized by performing more computations in place, instead of allocating specific buffers for each step of the computation pipeline. This will also be the subject of future developments.

\subsection{Profiling and benchmarking}
Holovibes supports multiple applications, including Doppler holography, static and dynamic optical coherence tomography (OCT) techniques (swept-source holographic OCT, full-field OCT and dynamic full-field OCT), holographic vibrometry, and digital holographic microscopy. Depending on the application type, reconstruction parameters, computer hardware, and camera configuration, the computational workload and system bottlenecks will vary.
For instance, with large image sizes, the GPU becomes the bottleneck—due to memory limitations when using large batches and time windows, or due to computation time when smaller frame packets are processed. Conversely, for small image sizes at high throughput (typical in Doppler holography), bottlenecks often arise from camera frame acquisition and data transfer rates. Extensive system profiling and benchmarking were conducted throughout development. Further profiling is planned to document each computational pipeline, analyze and compare workloads, identify bottlenecks, and guide future optimization.

\section{Conclusion}

We have developed a ultrahigh-speed digital hologram rendering software that enables real-time visualization of computed images, with a primary focus on retinal imaging applications. Using Holovibes, we demonstrated that interferograms captured at tens of thousands of frames per second can be processed and rendered in real time on a commodity PC and GPU setup. This capability allowed us to observe and analyze retinal blood flow dynamics on the fly, extracting quantitative biomarkers such as blood velocity and arterial pulsatility that were previously obtainable only through offline processing. The Holovibes engine achieves this performance by heavily exploiting parallelization and data buffering at every stage of the pipeline. Each core task – high-speed frame acquisition, numerical propagation and Doppler computation, and data recording – is executed asynchronously in its own thread, all running concurrently. Thread-safe ring buffers (allocated in GPU memory) serve as FIFOs between stages, ensuring a continuous streaming workflow without race conditions or bottlenecks. A global state manager coordinates configuration settings across threads, allowing dynamic adjustments (e.g. focus depth, processing mode) without interrupting the data flow. This modular software design cleanly separates the front-end visualization, the back-end processing, and the shared state control, which proved effective for maintaining stability at extreme data rates.

Experimental results confirm that ultrahigh-speed digital hologram rendering and short-time Doppler analysis can be performed with zero frame loss and low latency. In sustained tests, Holovibes successfully propagated the angular spectrum of 256×256 interferograms and computed STFT-based image rendering on 512-frame batches at 71,400 frames per second, all in streaming mode. This was achieved with no dropped frames using state-of-the-art high-speed cameras, while simultaneously recording the raw data to disk. The overall system throughput (input data rate) represents a significant improvement over our prior real-time holography implementations and, to our knowledge, exceeds the performance of other existing digital holography systems in the literature. The maximum end-to-end latency between an interferogram’s acquisition and its holographic image display was measured to be on the order of 30 ms, which is well within real-time requirements for interactive visualization and aligns with video framerate display cycles. This low latency and high frame rate ensure that even rapid dynamics are captured and displayed with minimal lag, enabling intuitive real-time observation of fast phenomena.

These advances carry broad implications for high-speed imaging in both biomedical and industrial contexts. In ophthalmology, the ability to process interferometric data at tens of kHz frame rates will facilitate in vivo retinal microvascular angiography non-invasively with unprecedented temporal resolution. This opens the door to monitoring fast hemodynamic transients and subtle vascular events that were previously inaccessible, potentially improving diagnosis and treatment monitoring for retinal diseases. Beyond biomedicine, the core architecture of Holovibes can be generalized to other domains requiring real-time analysis of high-volume data streams. In industrial metrology and materials science, Holovibes could enable live holographic inspection of microscopic vibrations or rapid structural changes, providing feedback in scenarios such as material fatigue testing or MEMS device characterization. By dramatically reducing processing latency, our approach allows these experiments to move from post-processing to real-time interactive analysis, which can yield new insights and more responsive control of experimental conditions.

Despite the demonstrated capabilities, our system also highlights areas for future improvement. Data transfer bottlenecks remain a limiting factor – in our current implementation, the transfer of frames from host memory to the GPU (over PCIe) is the primary throughput limiter. This suggests that even higher input rates could be supported if this bottleneck is alleviated, for example through faster bus interfaces. Moreover, the present results were achieved on relatively small interferogram sizes (256×256 pixels); scaling up to larger sensor formats at high frame rates will require further optimization and possibly more powerful computing resources. There is also room to expand the computational toolkit within Holovibes – incorporating advanced image processing to enhance the system’s versatility, though care must be taken to maintain real-time performance.

Future directions for this work will focus on several key aspects:
\begin{itemize}
    \item Throughput optimization: We plan to address the host–GPU transfer bottleneck by exploring high-bandwidth interfaces (such as PCIe 5.0 or GPU direct I/O) and more efficient memory management strategies, which could push the real-time throughput beyond the current 71,400 fps. Optimizing the pipeline to make full use of upcoming ultra-high-speed camera outputs is a priority.
    \item Scalability and distributed computing: To handle larger hologram sizes or multi-camera systems, the software may be extended to support multi-GPU processing. Parallelizing the workload across multiple GPUs or nodes could maintain real-time rates for higher resolution data or 3D holographic tomography, building on concepts demonstrated in cluster-based holographic displays.
    \item Flexible pipeline and API: We are developing a more flexible, modular processing pipeline that can be reconfigured or extended via a public API. This will allow users to insert custom processing modules (e.g. filtering, motion compensation, machine learning inference) into the real-time stream with minimal overhead, fostering broader use of the platform in various applications.
\end{itemize}

By pursuing these improvements, we aim to keep Holovibes at the cutting edge of high-speed computational imaging. As an open-source project, Holovibes will continue to evolve with contributions from the community, incorporating the latest algorithms and best practices. We believe that the real-time, ultrahigh-throughput holographic imaging demonstrated in this work not only serves immediate needs in biomedical research but also acts as a blueprint for future systems in other fields. The convergence of advanced imaging sensors, high-performance computing, and clever software design is driving a new generation of instruments that can capture and make sense of the dynamic world in real time. This progress will ultimately broaden the impact of digital holography and computational imaging, enabling new scientific discoveries and practical innovations across disciplines.

\bibliographystyle{IEEEtran}


\end{document}